\documentclass{article}

\usepackage{arxiv}

\usepackage{cite}
\usepackage{amsmath,amssymb,amsfonts}
\usepackage{algorithm}
\usepackage{algorithmic}
\usepackage{graphicx}
\usepackage{subcaption}
\usepackage{romannum}
\usepackage{booktabs}
\usepackage{multirow}
\usepackage{textcomp}
\usepackage[dvipsnames,table,xcdraw]{xcolor}
\usepackage{lipsum}
\usepackage{listings}
\usepackage{todonotes}
\usepackage{hyperref}
\usepackage{gnuplottex}
\usepackage{rotating}

\def\BibTeX{{\rm B\kern-.05em{\sc i\kern-.025em b}\kern-.08em
    T\kern-.1667em\lower.7ex\hbox{E}\kern-.125emX}}

\definecolor{blue-main}{rgb}{0,0,1}
\definecolor{dkgreen}{rgb}{0,0.6,0}
\definecolor{gray}{rgb}{0.5,0.5,0.5}
\definecolor{mauve}{rgb}{0.58,0,0.82}

\title{
    Towards Distributed Software Resilience in Asynchronous Many-Task Programming Models \\
    \begin{large}
    SNL Report Number: SAND2020-11278 C
    \end{large}
}

\author{
 Nikunj Gupta \\
  Dept. of CSE\\
  Indian Institute of Technology\\
   Roorkee, India\\
  \texttt{gnikunj@cct.lsu.edu} \\
   \And
 Jackson R. Mayo \\
  Sandia National Laboratories\\
  Livermore, California, USA\\
  \texttt{jmayo@sandia.gov} \\
  \And
 Adrian S. Lemoine \\
  AMD Inc.\\
  Austin, USA\\
  \texttt{adrian.lemoine@amd.com} \\
  \And
 Hartmut Kaiser \\
  Center for Computation Technology\\
  Louisiana State University\\
  Baton Rouge, USA \\
  \texttt{hkaiser@cct.lsu.edu} \\
}

\begin{document}

% code customization
\definecolor{codegreen}{rgb}{0,0.5,0}
\definecolor{codegray}{rgb}{0.5,0.5,0.5}
\definecolor{codered}{rgb}{0.75, 0.3 ,0.1}
 
\lstdefinestyle{mystyle}{
    commentstyle=\color{codegreen},
    keywordstyle=\color{codered},
    numberstyle=\tiny\color{codegray},
    basicstyle=\ttfamily\footnotesize,
    breakatwhitespace=false,         
    breaklines=true,                 
    captionpos=b,                    
    keepspaces=true,                 
    numbers=left,                    
    numbersep=8pt,                  
    showspaces=false,                
    showstringspaces=false,
    otherkeywords ={hpxr::async_replay, hpxr::async_replay_validate, hpxr::dataflow_replay, hpxr::dataflow_replay, hpxr::dataflow_replicate, hpxr::dataflow_replicate_validate, hpxr::dataflow_replicate_vote, hpxr::dataflow_replicate_vote_validate, hpxr::async_replicate, hpxr::async_replicate_validate, hpxr::async_replicate_vote, hpxr::async_replicate_vote_validate},
    showtabs=false,                  
    tabsize=2
}

\lstset{style=mystyle}

\maketitle
\begin{abstract}
Exceptions and errors occurring within mission critical applications due to hardware failures have a high cost. With the emerging Next Generation Platforms (NGPs), the rate of hardware failures will likely increase. Therefore, designing our applications to be resilient is a critical concern in order to retain the reliability of results while meeting the constraints on power budgets. In this paper, we discuss software resilience in AMTs at both local and distributed scale. We choose HPX to prototype our resiliency designs. We implement two resiliency APIs that we expose to the application developers, namely task replication and task replay. Task replication repeats a task n-times and executes them asynchronously. Task replay reschedules a task up to n-times until a valid output is returned. Furthermore, we expose algorithm based fault tolerance (ABFT) using user provided predicates (e.g., checksums) to validate the returned results. We benchmark the resiliency scheme for both synthetic and real world applications at local and distributed scale and show that most of the added execution time arises from the replay, replication or data movement of the tasks and not the boilerplate code added to achieve resilience.
\end{abstract}

% keywords can be removed

\section{Introduction}
The DOE Office of Science Exascale Computing Project (ECP)~\cite{bergman2008exascale} outlines the next milestones in the supercomputing domain. The target computing systems under the project will deliver 10x performance while keeping the power budget under 30
megawatts. With such large machines, the need to make applications resilient has become paramount. The benefits of adding resiliency to mission critical and scientific applications, includes the reduced cost of restarting the failed simulation both in terms of time and power.

Most of the current implementation of resiliency at the software level makes use of Coordinated Checkpoint and Restart (C/R)~\cite{daly2006higher, duell2005design, li1994low, moody2010design, plank1998diskless, roman2002survey}. This technique of resiliency generates a consistent global snapshot, also called a checkpoint. Generating snapshots involves global communication and coordination and is achieved by synchronizing all running processes. The generated checkpoint is then stored in some form of persistent storage. On failure detection, the runtime initiates a global rollback to the most recent previously saved checkpoint. This involves aborting all running processes, rolling them back to the previous state and restarting them.

In its current form, Coordinated C/R is excessively expensive on extreme-scale systems. This is due to the high overhead costs of global rollback followed by global restart. Adding to these overheads are the significant overheads of global I/O access. In many cases, millions of processes have to respond to a local process failure, which leads to heavy loss of useful CPU computation cycles and leads to a significant performance penalty. This was observed when node level resiliency was implemented in a production application running on Titan system at Oak Ridge National Laboratory~\cite{gamell2014exploring}. The overheads of resiliency had a significant impact on performance as the overheads of C/R were 20-30\% of the total execution time.

Emerging resilience techniques, such as Uncoordinated C/R~\cite{guermouche2011uncoordinated} and Local Failure Local Recovery (LFLR)~\cite{teranishi2014toward}, attempt to mitigate some of the overheads of coordinated C/R by eliminating the requirement of aborting all running processes and restarting. However, these techniques are based on assumptions exclusive to Single Program Multiple Data (SPMD) model, i.e., the same program execution across all running processes. Asynchronous Many-Task (AMT) execution models provide similar resilience techniques without any of these assumptions.

In this paper, we explore the implementation of resiliency techniques in AMT Runtime Systems. The design presented is general and can be applied to other AMTs. For prototyping software resilience, we chose HPX~\cite{Kaiser2020} as it exposes a standards conforming API which is easy to understand and adopt. AMTs replace the bulk-synchronous MPI model with fine-grained tasks and explicit task dependencies. They rely on a runtime system to schedule the tasks and manage their synchronization. In an AMT model, a program can be seen as a flow of data which is processed by tasks, each task executing a distinct kernel. Failures within a program are naught more than a manifestation of a failed task, which can be identified as a local point of failure. This significantly simplifies the implementations of a resilient interface. The research presented in this paper focuses on errors based on silent data corruptions (SDC), i.e., errors arising from unexpected aberrations in data or compute. We also consider memory bit flips in our testing. These error types are not detected by the operating system and pose a serious risk in exascale computing. To the best of our knowledge, this is the first work that discusses AMT resiliency on both local and distributed scale with further discussions on distributed design, implementation, overhead evaluations, and drawbacks. The paper makes the following key contributions:
\begin{enumerate}
    \item Design and implementation of local and distributed resilience techniques with C++ standards conforming APIs.
    \item Prototyping various composable resilience APIs and evaluating their performance.
    \item Execution of resilient and non-resilient tasks in HPX.
\end{enumerate}

\section{Related Work}

Software based resilience for SPMD programs has been well studied and explored including but not limited to coordinated checkpoint and restart (C/R). MPI-ULFM~\cite{fgcs20LosadaGMBBT} provides a good summary of ULFM solution to MPI applications for exascale applications. MPI\_Reinit~\cite{DBLP:journals/concurrency/ChakrabortyLEMP20,DBLP:conf/supercomputer/GeorgakoudisGL20} has also been proposed which directly accesses the resource manager of the cluster systems for quick online application recovery. MPI fault tolerance using the charm++ backend has also been studied and proved to be resilient~\cite{taskFTMPI}.

Enabling resilience in AMT execution models has not been well studied despite the fact that the AMT paradigm facilitates an easier implementation. Subasi \textit{et al} ~\cite{subasi2015nanocheckpoints, subasi2016runtime, subasi2017designing} have discussed a combination of task replay and replicate with C/R for task-parallel runtime, OmpSs. For task replication, they suggested to defer launch of the third replica until duplicated tasks report a failure. This differs from our implementation, as we replicate the tasks and do not defer the launch of any task. For task replay, they depend on the errors triggered by the operating system. This approach, thus, assumes reliable failure detection support by the operating system, which is not always available. We also found that automatic global checkpointing has been explored within the Kokkos ecosystem~\cite{edwards2014kokkos,miles2019software}.

Similar resilience work has recently been explored with HClib~\cite{paul2019enabling}. The work, however, is based on on-node resiliency. The research we present is generalized and unified, i.e., the APIs exposed allow for both local and distributed resilience. Furthermore, we provide a finer control over the APIs by introducing multiple variants of a single resilience API. Cao et al.~\cite{7161563} explored resilience in distributed runtime, but they consider static task graphs, and lacks general applicability. Our work differs as our resilience APIs work dynamically.  The work we present in this paper is a direct extension of our previous resilience work~\cite{gupta2020implementing}.

\subsection{HPX}

HPX~\cite{Kaiser2020,heller2013application,heller2013using,kaiser2015higher,heller2016closing,kaiser2014hpx} is a C++ standard library for distributed and parallel programming built on top of an asynchronous many-task (AMT) runtime system. Such AMT runtimes may provide a means for helping programming models to fully exploit available parallelism on complex emerging HPC architectures. The HPX programming model includes the following essential components: {\it (1)} an ISO C++ standard conforming API that enables wait-free asynchronous parallel programming, including futures, channels, and other primitives that enable asynchronous operation~\cite{heller2019harnessing}; {\it (2)} an active global address space (AGAS) that supports load balancing via object migration~\cite{amini2019agas}; {\it (3)} an active-message networking layer that ships functions to the objects they operate on~\cite{biddiscombe2017zero}; {\it (4)} work-stealing lightweight task scheduler that enables finer-grained parallelization and synchronization~\cite{kaiser2009parallex,grubel_2015}. 

\section{AMT Resiliency Design}

% \subsection{AMT Resiliency Design}

In an AMT programming model, the program execution can be visualized as a directed acyclic graph (DAG) with each graph node as a task that depends on its parent node(s), and whose children depend on its execution. Furthermore, tasks themselves do not involve internal synchronization, i.e., a task that gets scheduled runs to completion without scheduler intervention. A task may further invoke asynchronous or synchronous tasks which can then be visualized as a DAG. This makes the task boundary a prospective place for additional resiliency checks for error detection and corrections. We prototype resiliency APIs around task boundaries for the same reasons.

Before replaying/replicating a task, we need to ensure the correctness of the input data and the global state. AMTs in general discourage the use of global variables for communication and promotes built-in constructs. Therefore in our prototype APIs, we assume that a task does not change the state of global variables (provided there are any). Furthermore, we do not store the input data either to memory or disk. I/O operations on input argument data of large sizes is a major bottleneck, and can increase the execution time by multiple folds. Instead we keep a decayed copy~\footnote{\url{https://en.cppreference.com/w/cpp/types/decay}} of the input. We assume that the task does not change the input data arguments, or the arguments changed are orthogonal to the inputs for the algorithms carried within the task.

We extend the basic local resilience to a distributed scale. Here, we assume that the network is reliable, i.e., any data transferred through the wire is not altered or prone to corruption. One can extend the proposed facilities to be network resilient by generating a hash function over the input data and transferring it with the input data over the wire. If the hashes computed at the receiving site match the sent hashes, the algorithm can proceed, otherwise, the recipient can request a new set of input data. The implementation comes with an additional hash computation overhead and makes our prototyped APIs more complex. Therefore, discussing its implementation within the prototyped APIs is out of the scope of this research.

Another point of interest is transferring data over the wire. In a local resilience scheme, an application can send large data objects as constant reference to the task. The task then executes an algorithm that takes the input data and generates a result. At a distributed scale, such a scheme is not possible. There can be algorithm specific workarounds to this issue, but a generalized solution to mitigate high data transfer overheads remain an open question.

Finally, for distributed resilience, the functions are initiated at the locality (i.e. node) that calls the distributed resilience, i.e., the locality that calls distributed resilience APIs manages failures and validation of checksums. Another implementation of distributed resilience could be offloading the whole facility to the localities they are invoked on. The latter implementation does relieve network communication costs by invoking the validation and consensus functions on the same locality as the task. We chose the former implementation as a user can invoke a task on another locality and use local resilience variations to achieve the same effect as the latter implementation.

In this paper, we focus mainly on errors that go undetected by the operating system and require special handling at the application level. Such errors include silent data corruptions and memory bit flips. Starting now, we categorize these errors as ``failure points'' or simply as ``failures''. From the AMT design characteristics defined above, we can conclude that a ``failure'' is a manifestation of a failing task. Failing tasks can occur either through exceptions (e.g., arising from memory bit flips) or through failing validation checks, such as in algorithm based fault tolerance using checksums.

\section{Implementation Details}

Using the resiliency design described above, we find HPX a suitable platform to perform experiments with resiliency APIs. We present two different ways to expose resiliency capabilities to the user, namely task replay and task replicate. This section discusses the implementation details of these APIs and the required changes to the code.

HPX introduces parallelism using parallel execution policies. HPX executors sit on top of these execution policies. An executor describes how the execution policy should be invoked. An executor is then passed to various HPX facilities to achieve the desired behavior. For example, \texttt{block\_executor} provides NUMA aware execution of a parallel algorithm when used in \texttt{parallel\_for} loop. We begin by first creating resilience executors for HPX. A resilience executor takes in an execution policy and the number of replays or replications. These executors can then be passed to HPX's parallel algorithms to achieve resilience. While these executors will work for standard \texttt{async} and \texttt{dataflow} facilities in HPX, we decided to create separate resilience wrappers around them as well. These wrappers provide a finer control over the APIs and include several variants for each resilience type, i.e., replay and replicate.

\subsection{Task Replay}
\label{sub:replay}

Task Replay is analogous to the Checkpoint/Restart mechanism found in conventional execution models. The key difference is localized fault detection. When the runtime detects a failure it replays the failing task as opposed to complete roll back of the entire program. Our prototype APIs revolve around two key additional arguments to \texttt{async}. The first additional argument is \textit{N}, i.e., the number of times a task must be automatically replayed before giving up on error correction. The second argument is a \textit{validate} function that can be used for algorithm based fault tolerance. Based on these two arguments, we designed two variations of task replay namely:

(\romannum{1}) \textbf{Async and Dataflow Replay:} This version of task replay will catch user defined exceptions, and automatically reschedule the task up to {\it N} times if an exception is caught, before re-throwing the last caught exception.

(\romannum{2}) \textbf{Async and Dataflow Replay Validate:} This version of task replay extends (\romannum{1}) and adds a \textit{validate} function. This user defined function can be used to add algorithm based fault tolerance. For instance, a user may decide to compare the returned result using checksums to identify if any SDC based errors occurred. Here, a task is replayed until either no failures are encountered by both the task and the validation function, or number of replays are exhausted.

HPX allows distributed task execution using actions. User wraps a function to an action (a serializable entity) which is then passed to \texttt{async} with the locality where the task should be invoked. To extend the current local resilience facilities to a distributed setting, we require one additional argument, namely a list of localities where the task is required to be invoked. The order of localities dictates the order in which a task will be replayed. Furthermore, the user needs to pass an action to the API instead of the function name.

\subsection{Task Replicate}\label{AA}
\label{sub:replicate}

Task Replicate is designed to provide reliability enhancements by replicating a set of tasks and evaluating their results to determine a consensus among them. This technique is most effective in situations where only a few tasks are executing in the critical path of the DAG leaving the system underutilized.
% there are few tasks in the critical path of the DAG which leaves the system underutilized. 
However, the drawback of this method is the additional computational cost incurred by replicating a task multiple times. Our prototype APIs revolve around three key additional arguments to \texttt{async}. The first two additional arguments are the same as described for Task Replay. The third additional argument is a feature specific to task replicate, i.e., a consensus function. This function is essential for cases where multiple returned results pass the validation phase. Using this function, the user can implement their own consensus function that returns the result required by the user. Based on these three arguments, we designed four variations of task replication, namely:

(\romannum{1}) \textbf{Async and Dataflow Replicate:} %This is the most basic 
%implementation of the task replication. 
This API returns the first  result that runs without failures.

(\romannum{2}) \textbf{Async and Dataflow Replicate Validate:} This
API additionally takes a function that validates the individual results. 
It returns the first result that is positively validated. \textit{validate} API works similar to the one described in task replay.

(\romannum{3}) \textbf{Async and Dataflow Replicate Vote:} This API 
adds a \textit{consensus} function to the basic replicate function. Many hardware or 
software failures are silent errors that do not interrupt the program flow.
%In order to detect errors of this kind, it is necessary to run the task several 
%times and 
The API determines the ``correct'' result by using the voting function allowing
to build a consensus. 
%In order to determine which return value is ``correct", the API
%allows the user to provide a custom consensus function to properly form a
%consensus. This voting function then returns the ``correct"
%answer.

(\romannum{4}) \textbf{Async and Dataflow Replicate Vote Validate:} This 
combines the features of the previously discussed replicate APIs. Replicate 
vote validate allows a user to provide a \textit{validation} function and
a \textit{voting} function to filter results.

We extend the above APIs to a distributed scale with a similar approach as described for Task Replay. The current implementation does not allow for duplicate tasks to get cancelled in an event where one of the tasks returns a valid output. Future work will include cancellable tasks which will signal all duplicate tasks to deque and not execute.

\subsection{Usage}

Listing~\ref{listing:usage} describes the required changes to the code. Both local and distributed resilience APIs resemble closely their non-resilient counterparts, with the additional arguments described in ~\ref{sub:replay} and ~\ref{sub:replicate}.

For distributed resilience, we currently demand the user to provide a set of localities on which a task will be invoked either concurrently (resilience replicate) or in a round robin manner (resilience replay). In the future, we wish to replace this facility by taking a user provided load balancing executor. The load balancer will then schedule the task on a locality with starving/least loaded processors.

\begin{lstlisting}[frame=single, language=C++, caption={Required code changes to utilize the resilience variations.}, label={listing:usage}]
// Our task
int univ_ans() { return 42; }

// Defining an action over the task
HPX_PLAIN_ACTION(univ_ans, universal_action);
// Our validate function
bool validate(int res) { return res == 42; }

// Non-resilient API
async(univ_ans); // Local
universal_action ac; // Action object
async(ac, find_here()) // Distributed

// Resilient Local API
async_replay_validate(3, validate, univ_ans);
async_replicate_validate(3, validate, univ_ans);

// Resilient Distributed API
async_replay_validate(ids, validate, ac);
async_replicate_validate(ids, validate, ac);
\end{lstlisting}

\section{Benchmarks}
\label{sec:benchmarks}

This section discusses synthetic and real world applications that were used to measure the overheads of our resilience APIs. The source code and the tests as described below are currently under review post which they'll be merged into HPX~\footnote{\url{https://github.com/STEllAR-GROUP/hpx/pull/4858}}. It can be found as an independent module named `resiliency' under the lib directory.

\subsection{Synthetic Local Workloads}

This synthetic benchmark is written to imitate an actual application, while providing us the means to change parameters, namely grain-size and task count. This allows us to accurately measure the implementation overheads of our resilience APIs while changing the parameters to suit a more realistic application.

HPX works best when the grain-size of a task is greater than 200$\mu$s. Therefore, we test the resiliency APIs with a grain-size of 200$\mu$s. To allow for sufficient parallelism across all 48 cores on the processor we test on, we invoke a total of one million tasks. This scenario is very similar to an actual application written in HPX.

\subsection{Synthetic Distributed Workloads}

This synthetic benchmark is written to imitate an actual distributed HPX application, while providing us the means to change parameters, namely number of actions invoked, number of tasks invoked per actions, and the grain size of these invoked tasks. Having control over the above parameters allow us to measure distributed resilience API overheads in a more real world application scenario.

For this benchmark, we invoke a total of 25,000 actions on different localities with each action triggering another thousand tasks to run on that locality. Each of these tasks is 500$\mu$s in size.

\subsection{1D Stencil Local}
\label{subsection:1d_stencil_local}

For 1D stencil, we port the 1D stencil application from HPX benchmark suite with two key changes. Firstly, we add the feature to advance multiple time steps in each iteration of the stencil. This allows us to do more work per iteration with significantly less communication overhead. It is achieved by reading an extended ``ghost region'' of data values from each neighbor. Secondly, we add physics based checksums to verify the validity of the output. The benchmark solves a linear advection equation. The task decomposition, Lax-Wendroff stencil, and checksum operations are as described in previous work~\cite{paul2019enabling}. We use HPX \texttt{dataflow} to implement the benchmark. \texttt{dataflow} allows to synchronize between multiple tasks, in our case three. Each task waits on the current subdomain, and the neighboring left and right subdomains.

We run the benchmark with two cases that we call 1D stencil case A and 1D stencil case B. Case A uses 384 subdomains each with 8,000 data points while case B uses 192 subdomains each with 16,000 data points. Both cases runs for 4,096 iterations with 256 time steps per iteration.

\subsection{1D Stencil Distributed}
\label{subsection:stencil_distributed}

For distributed resilience performance, 1D stencil local is extended further to support distributed functionalities. This is achieved by calling the HPX initialization function on all localities and introducing HPX channels to allow for communication between localities. The implementation follows a traditional lockstep model instead of the dataflow approach. This is done to keep the implementation complexity minimal. Furthermore, we implement the benchmark with a mix of local and distributed resilience. Adding local resilience allows us to try a couple of times before giving up in cases where a node is faulty. This ensures that large grids are not moved over the wire unless there is an absolute need for it. Also, it shows how easily a user can wrap local resilience with distributed resilience facilities. 

The benchmark is run for the cases described in ~\ref{subsection:1d_stencil_local} with one key difference. The iterations are reduced to 512 and the steps per iteration is increased to 2048. This is done to increase the grain size of the task, which helps in overlapping the network latencies.

\subsection{Faults and Errors}
\label{subsection:errors}

The errors that we inject within the applications are completely artificial and not a reflection of any computational or memory failure. We introduce errors by utilizing C++ exception facility. Error probability as described in the plots in Section~\ref{sec:results} is the probability of a failing task. This means that a failed task that is replaying itself has the same probability of failure. This helps us to simulate a real world experience.

We define a faulty node as a node with ten times the probability of SDC errors when compared to a standard node. The rational behind selecting the number ten is to signify that a faulty node has a significantly higher probability of failing when compared to standard nodes. In reality the number could be higher or lower. For testing purposes, we found ten to be large enough to highlight a faulty node. For instance, if a benchmark is provided with an error rate of 5\%, a normal node will simulate it at an error rate of 5\%, while, a faulty node will simulate it at an error rate of 50\%. We render a node faulty using a list of ids that are passed to our benchmarks. If a node rank belongs to the list of faulty nodes, it marks itself as faulty. These are randomly assigned localities to simulate real world experience.

\section{System Setup}

We use the newly established LONI cluster at LSU. Each node has two sockets, equipped with Intel Xeon Platinum 8260 CPU @ 2.40GHz, for a total of 24 cores per socket and 48 cores per node. The nodes are equipped with 382GB RAM each.

Table~\ref{tab:config} lists the versions of all the dependencies of HPX. All prototype APIs were designed and developed on the current master branch of HPX.

\begin{table}[htb]
  \centering
  \caption{HPX dependencies used for prototyping APIs}
  \label{tab:config}
  \begin{tabular}{ll|ll}
    \toprule
    % \textbf{Package Name}       & \textbf{Version}  \\ \hline
    % \midrule
    Compiler                         & gcc 9.3 & hwloc                       & 1.1              \\ \hline
    jemalloc                    & 5.2.1 & boost                       & 1.73               \\ \hline
    \bottomrule
  \end{tabular}
\end{table}

To ensure statistically relevant results, we run all the benchmarks three times. We report the least execution time reported in one of the three runs. Furthermore, we do not include the runtime initialization and shutdown costs in the measured execution time. Since the errors are probabilistic, it is understood that each run will have different error count. This is a design decision taken during the implementation of the benchmark to simulate a real world experience. To mitigate this issue, we rerun the benchmarks if the variance in execution times differ by more than one percent. This makes sure that outlier cases with significant differences in error counts are dealt with.

\section{Results}
\label{sec:results}

This section discusses the empirical results for the benchmarks described in Section~\ref{sec:benchmarks}.

\subsection{Synthetic Local Workloads}

\begin{figure}[tphb]
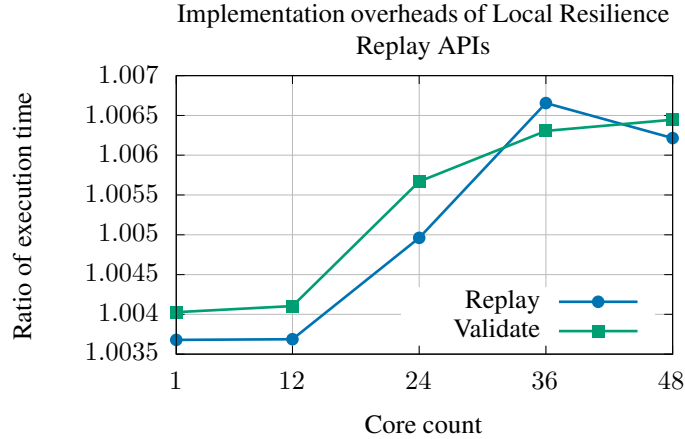
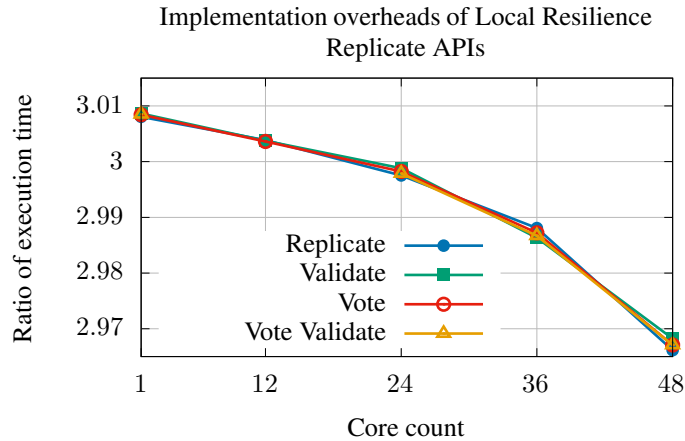

    \centering
    \begin{subfigure}{\linewidth}
        \centering
        \begin{gnuplot}[terminal=epslatex]
            set terminal epslatex size 3.8in,2.4in
            set ylabel 'Ratio of execution time'
            set xlabel 'Core count'
            
            set grid ytics lt 2 lc rgb "#bbbbbb"
            set grid xtics lt 2 lc rgb "#bbbbbb"
            
            set title '\shortstack{Implementation overheads of Local Resilience\\Replay APIs}'
            set key bottom right
            
            set xtics (1, 12, 24, 36, 48)
    
            plot \
            'plots/local/overheads_replay_200' using 1:($3/$2) title 'Replay' lw 4 lt 7 lc 6 ps 1.5 w lp, \
            'plots/local/overheads_replay_200' using 1:($4/$2) title 'Validate' lw 4 lt 5 lc 2 ps 1.5 w lp 
        \end{gnuplot}
        \caption{Ratio of execution time for resiliency replay and replay validate to pure async. The application is run with a million tasks, each with a grain-size of 200$\mu$s.}
        \label{fig:overheads_replay_local}
    \end{subfigure}
    \begin{subfigure}{\linewidth}
        \centering
        \begin{gnuplot}[terminal=epslatex]
            set terminal epslatex size 3.8in,2.4in
            set ylabel 'Ratio of execution time'
            set xlabel 'Core count'
            
            set yrange[2.965:3.015]
            
            set grid ytics lt 2 lc rgb "#bbbbbb"
            set grid xtics lt 2 lc rgb "#bbbbbb"
            
            set title '\shortstack{Implementation overheads of Local Resilience\\Replicate APIs}'
            set key bottom left
            
            set xtics (1, 12, 24, 36, 48)
    
            plot \
            'plots/local/overheads_replicate_200' using 1:($3/$2) title 'Replicate' lw 4 lt 7 lc 6 ps 1.5 w lp, \
            'plots/local/overheads_replicate_200' using 1:($4/$2) title 'Validate' lw 4 lt 5 lc 2 ps 1.5 w lp, \
            'plots/local/overheads_replicate_200' using 1:($5/$2) title 'Vote' lw 4 lt 6 lc 7 ps 1.5 w lp, \
            'plots/local/overheads_replicate_200' using 1:($6/$2) title 'Vote Validate' lw 4 lt 8 lc 4 ps 1.5 w lp 
        \end{gnuplot}
        \caption{Ratio of execution time for resiliency replicate and variants to pure async. The application is run with a million tasks, each with a grain-size of 200$\mu$s.}
        \label{fig:overheads_replicate_local}
    \end{subfigure}
    \caption{Implementation overheads of resilience local APIs.}
    \label{fig:overheads_local}
\end{figure}

Figure~\ref{fig:overheads_replay_local} and Figure~\ref{fig:overheads_replicate_local} succinctly represents the implementation overheads of adding resilience. For resilience replay variations, the observed overheads are minimal and less than 1\% of extra execution time. The added execution time can be considered noise for most practical applications. This is an expected behavior as we rely on the advantages of an AMT model, namely task communication using channels and other AMT native primitives instead of relying on global variables. This allows us to keep a decayed copy instead of keeping a sane global state during each resilience API invocation.

\begin{figure}[thbp]
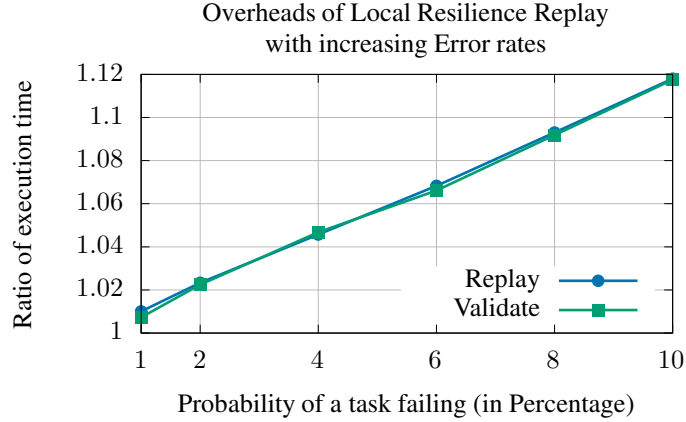
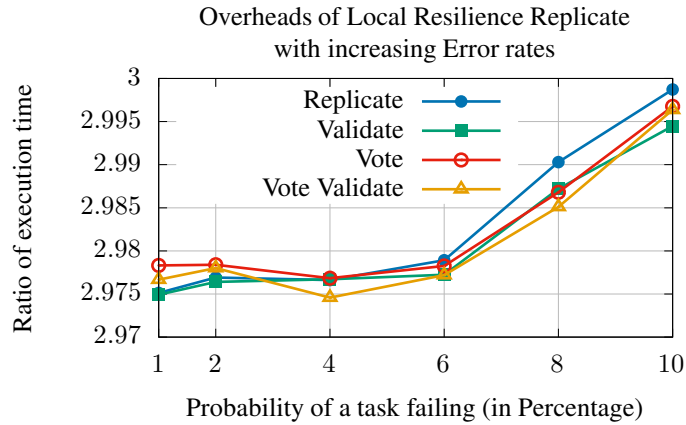

    \centering
    \begin{subfigure}{\linewidth}
        \centering
        \begin{gnuplot}[terminal=epslatex]
            set terminal epslatex size 3.8in,2.3in
            set ylabel 'Ratio of execution time'
            set xlabel 'Probability of a task failing (in Percentage)'
            
            set grid ytics lt 2 lc rgb "#bbbbbb"
            set grid xtics lt 2 lc rgb "#bbbbbb"
            
            set title '\shortstack{Overheads of Local Resilience Replay\\with increasing Error rates}'
            set key bottom right
            
            set xtics (1,2,4,6,8,10)
    
            plot \
            'plots/local/errors_replay_200' using 1:($3/$2) title 'Replay' lw 4 lt 7 lc 6 ps 1.5 w lp, \
            'plots/local/errors_replay_200' using 1:($4/$2) title 'Validate' lw 4 lt 5 lc 2 ps 1.5 w lp
        \end{gnuplot}
        \caption{Ratio of execution times for resiliency replay and replay validate to pure async. The application is run with a million tasks, each with a grain-size of 200$\mu$s, and utilizes all 48 cores.}
        \label{fig:errors_replay_local}
    \end{subfigure}
    
    \begin{subfigure}{\linewidth}
        \centering
        \begin{gnuplot}[terminal=epslatex]
            set terminal epslatex size 3.8in,2.3in
            set ylabel 'Ratio of execution time'
            set xlabel 'Probability of a task failing (in Percentage)'
            
            set yrange[2.97:3]
            
            set grid ytics lt 2 lc rgb "#bbbbbb"
            set grid xtics lt 2 lc rgb "#bbbbbb"
            
            set title '\shortstack{Overheads of Local Resilience Replicate\\with increasing Error rates}'
            set key top left
            
            set xtics (1,2,4,6,8,10)
    
            plot \
            'plots/local/errors_replicate_200' using 1:($3/$2) title 'Replicate' lw 4 lt 7 lc 6 ps 1.5 w lp, \
            'plots/local/errors_replicate_200' using 1:($4/$2) title 'Validate' lw 4 lt 5 lc 2 ps 1.5 w lp, \
            'plots/local/errors_replicate_200' using 1:($5/$2) title 'Vote' lw 4 lt 6 lc 7 ps 1.5 w lp, \
            'plots/local/errors_replicate_200' using 1:($6/$2) title 'Vote Validate' lw 4 lt 8 lc 4 ps 1.5 w lp 
        \end{gnuplot}
        \caption{Ratio of execution times for resiliency replicate and variants to pure async. The application is run with a million tasks, each with a grain-size of 200$\mu$s, and utilizes all 48 cores.}
        \label{fig:errors_replicate_local}
    \end{subfigure}
    \caption{Overheads of resilience local APIs with increasing error rates for a task grain-size of 200$\mu$s.}
    \label{fig:overheads_local}
\end{figure}

For resilience replicate variations, the execution time correlates to the number of replicates. In our case, three replicates causes the execution time to become three fold. While our synthetic benchmark does not attempt to take advantage of the caches, we see inherent cache benefits. Furthermore, the implementation overheads themselves are negligible for replicate variations as well. The minor differences in overheads between resilient variants arises from the underlying implementation, some requiring more boilerplate code than others.

Figure~\ref{fig:errors_replay_local} illustrates the added execution time that one can expect with increasing software faults. When errors are encountered, the resilient logic is activated and behaves as specified. For cases with low probability of failures, we see that amortized overheads of resilience replay APIs are still small enough to be hidden by system noise. Given that the probability of failure within a machine will not be more than a percent in most cases, it is safe to assume that async replay introduces no measurable overheads for applications utilizing the feature.
% Taken together, the presented results indicate that these resilient features will not incur any meaningful execution time costs.

For replicate variants, we do not see any changing behavior with increasing error rates. This is expected as the replicates are overheads themselves. Replication involves costly overheads, and it is recommended to be used in portions of code which are starving for work (i.e., insufficient parallelism) or for critical portions of code.

\subsection{Synthetic Distributed Workloads}

\begin{figure}[htbp]
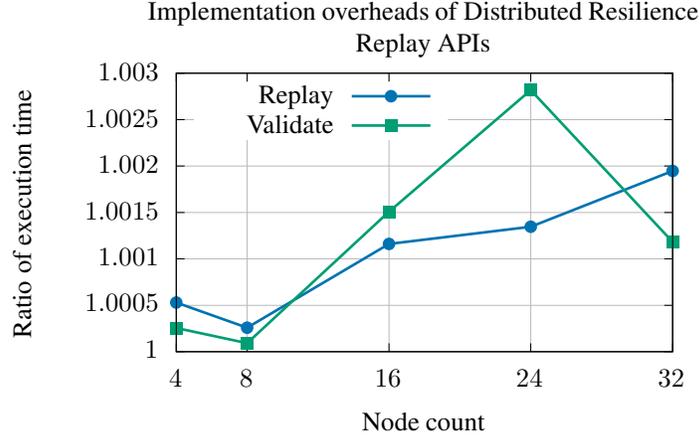
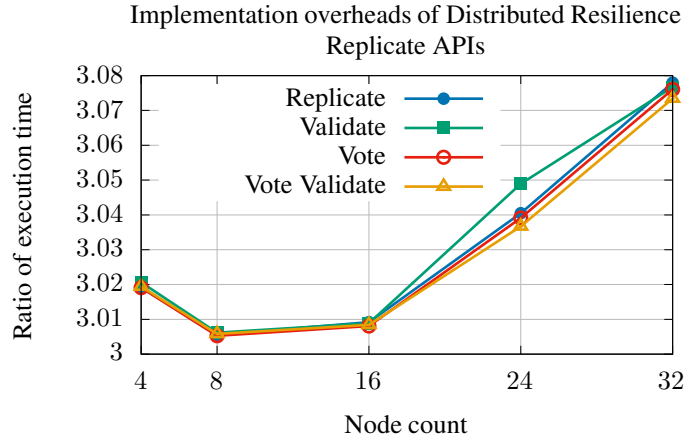

    \begin{subfigure}{\linewidth}
        \centering
        \begin{gnuplot}[terminal=epslatex]
            set terminal epslatex size 3.8in,2.4in
            set ylabel 'Ratio of execution time'
            set xlabel 'Node count'
            
            set yrange[1:1.003]
            
            set grid ytics lt 2 lc rgb "#bbbbbb"
            set grid xtics lt 2 lc rgb "#bbbbbb"
            
            set title '\shortstack{Implementation overheads of Distributed Resilience\\Replay APIs}'
            set key top left
            
            set xtics (4,8,16,24,32)
    
            plot \
            'plots/distributed/overheads_replay_500' using 1:($3/$2) title 'Replay' lw 4 lt 7 lc 6 ps 1.5 w lp, \
            'plots/distributed/overheads_replay_500' using 1:($4/$2) title 'Validate' lw 4 lt 5 lc 2 ps 1.5 w lp 
        \end{gnuplot}
        \caption{Ratio of execution times taken for resiliency replay and replay validate API compared to pure async. The application is run with 25,000 actions, each action invoking a thousand tasks with task grain-size of 500$\mu$s.}
        \label{fig:overheads_replay_distributed}
    \end{subfigure}
    
    \begin{subfigure}{\linewidth}
        \centering
        \begin{gnuplot}[terminal=epslatex]
            set terminal epslatex size 3.8in,2.4in
            set ylabel 'Ratio of execution time'
            set xlabel 'Node count'
            
            set grid ytics lt 2 lc rgb "#bbbbbb"
            set grid xtics lt 2 lc rgb "#bbbbbb"
            
            set title '\shortstack{Implementation overheads of Distributed Resilience\\Replicate APIs}'
            set key top left
            
            set xtics (4,8,16,24,32)
    
            plot \
            'plots/distributed/overheads_replicate_500' using 1:($3/$2) title 'Replicate' lw 4 lt 7 lc 6 ps 1.5 w lp, \
            'plots/distributed/overheads_replicate_500' using 1:($4/$2) title 'Validate' lw 4 lt 5 lc 2 ps 1.5 w lp, \
            'plots/distributed/overheads_replicate_500' using 1:($5/$2) title 'Vote' lw 4 lt 6 lc 7 ps 1.5 w lp, \
            'plots/distributed/overheads_replicate_500' using 1:($6/$2) title 'Vote Validate' lw 4 lt 8 lc 4 ps 1.5 w lp 
        \end{gnuplot}
        \caption{Ratio of execution times taken for resiliency replicate APIs compared to pure async. The application is run with 25,000 actions, each action invoking a thousand tasks with task grain-size of 500$\mu$s.}
        \label{fig:overheads_replicate_distributed}
    \end{subfigure}
    \caption{Implementation overheads of resilience distributed APIs.}
    \label{fig:overheads_distributed}
\end{figure}

Figure~\ref{fig:overheads_replay_distributed} and Figure~\ref{fig:overheads_replicate_distributed} describe the implementation overheads of adding resilience. For resilience replay and replicate variations, the observed overheads are very similar to the local counterparts. The added execution time is attributed to extra communication costs that distributed variations have to bear.

The benchmark does not move data from one locality to the other, so the communication overheads are minimal. An application that relies on heavy data transfers will observe significant overheads due to network bandwidth bottlenecks. It is recommended that such applications either use local resilience variations, or transfer data once every few iterations with optimal network latency hiding techniques.

\begin{figure}[htbp]
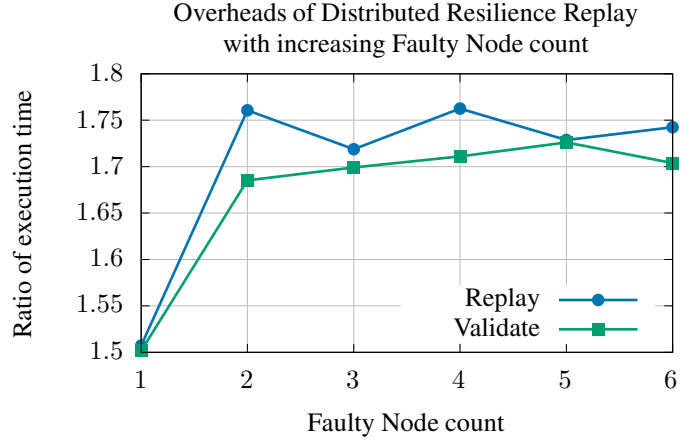
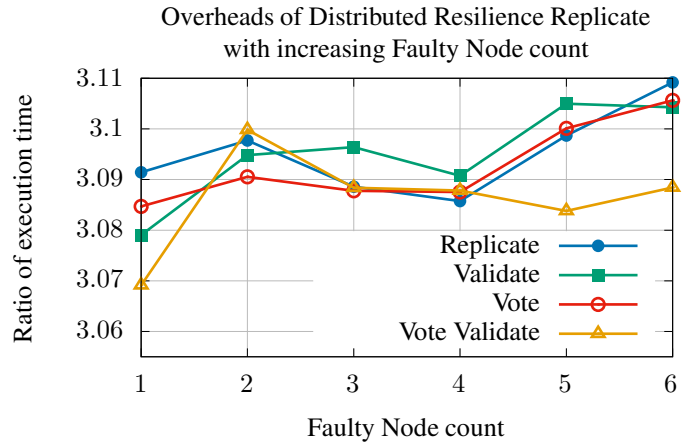

    \begin{subfigure}{\linewidth}
        \centering
        \begin{gnuplot}[terminal=epslatex]
            set terminal epslatex size 3.8in,2.4in
            set ylabel 'Ratio of execution time'
            set xlabel 'Faulty Node count'
            
            set grid ytics lt 2 lc rgb "#bbbbbb"
            set grid xtics lt 2 lc rgb "#bbbbbb"
            
            set title '\shortstack{Overheads of Distributed Resilience Replay\\with increasing Faulty Node count}'
            set key bottom right
            
            set xtics (1,2,3,4,5,6)
    
            plot \
            'plots/distributed/faults_replay_500' using 1:($3/$2) title 'Replay' lw 4 lt 7 lc 6 ps 1.5 w lp, \
            'plots/distributed/faults_replay_500' using 1:($4/$2) title 'Validate' lw 4 lt 5 lc 2 ps 1.5 w lp 
        \end{gnuplot}
        \caption{Ratio of execution times for resiliency replay and replay validate to pure async. The application is run with 25,000 actions, each action invoking a thousand tasks with task grain-size of 500$\mu$s.}
        \label{fig:faults_replay_distributed}
    \end{subfigure}
    
    \begin{subfigure}{\linewidth}
        \centering
        \begin{gnuplot}[terminal=epslatex]
            set terminal epslatex size 3.8in,2.4in
            set ylabel 'Ratio of execution time'
            set xlabel 'Faulty Node count'
            
            set yrange[3.055:3.11]
            
            set grid ytics lt 2 lc rgb "#bbbbbb"
            set grid xtics lt 2 lc rgb "#bbbbbb"
            
            set title '\shortstack{Overheads of Distributed Resilience Replicate\\with increasing Faulty Node count}'
            set key bottom right
            
            set xtics (1,2,3,4,5,6)
    
            plot \
            'plots/distributed/faults_replicate_500' using 1:($3/$2) title 'Replicate' lw 4 lt 7 lc 6 ps 1.5 w lp, \
            'plots/distributed/faults_replicate_500' using 1:($4/$2) title 'Validate' lw 4 lt 5 lc 2 ps 1.5 w lp, \
            'plots/distributed/faults_replicate_500' using 1:($5/$2) title 'Vote' lw 4 lt 6 lc 7 ps 1.5 w lp, \
            'plots/distributed/faults_replicate_500' using 1:($6/$2) title 'Vote Validate' lw 4 lt 8 lc 4 ps 1.5 w lp 
        \end{gnuplot}
        \caption{Ratio of execution times for resiliency replicate APIs to pure async. The application is run with 25,000 actions, each action invoking a thousand tasks with task grain-size of 500$\mu$s.}
        \label{fig:faults_replicate_distributed}
    \end{subfigure}
    \caption{Overheads of resilience distributed APIs with increasing number of faulty nodes.}
    \label{fig:faults_distributed}
\end{figure}

\begin{figure}[bp]
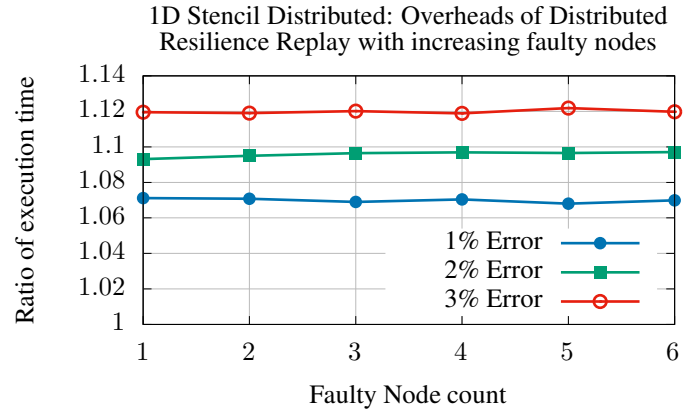

    \centering
    \begin{gnuplot}[terminal=epslatex]
        set terminal epslatex size 3.8in,2.25in
        set ylabel 'Ratio of execution time'
        set xlabel 'Faulty Node count'
        
        set grid ytics lt 2 lc rgb "#bbbbbb"
        set grid xtics lt 2 lc rgb "#bbbbbb"
        
        set title '\shortstack{1D Stencil Distributed: Overheads of Distributed\\Resilience Replay with increasing faulty nodes}'
        set key bottom right
        
        set yrange[1:1.14]
        
        set xtics (1,2,3,4,5,6)

        plot \
        'plots/1d_stencil_distributed/errors_384' using 1:($3/$2) title '1\% Error' lw 4 lt 7 lc 6 ps 1.5 w lp, \
        'plots/1d_stencil_distributed/errors_384' using 1:($4/$2) title '2\% Error' lw 4 lt 5 lc 2 ps 1.5 w lp, \ 
        'plots/1d_stencil_distributed/errors_384' using 1:($5/$2) title '3\% Error' lw 4 lt 6 lc 7 ps 1.5 w lp
    \end{gnuplot}
    \caption{Ratio of execution times for resiliency replay APIs to base non-resilient APIs with increasing faulty node count. The case works on 384 subdomains each with 8,000 data points and iterate over 512 iterations with 2048 time steps per iteration.}
    \label{fig:1d_stencil_distributed_errors}
\end{figure}

When exposed to faulty nodes (See Figure~\ref{fig:faults_distributed}), the execution time increases for the replay variations visibly up to 2 nodes, after which it flattens out. This is an expected behavior and can be explained by understanding the design of the benchmark. The benchmark is designed such that tasks are invoked on the next locality if it fails to execute as expected. This means that all failed tasks invoked on a faulty node execute on the next node by rank. This design was a conscious decision to showcase how a non-balanced distributed resilience API can cause significant rise in execution times. When a single node is faulty, some actions invoking on that locality are bound to fail (we describe faulty nodes with failure rate of 50\%, see Section~\ref{subsection:errors}). All these actions are then migrated to another node where they are executed. Meanwhile, other localities are starving for work. On increasing the number of faulty nodes, we see a flattening behavior as less localities are starving now coupled with parallel execution of actions on various localities. An application leveraging asynchrony coupled with sufficient parallelism will observe a noticeably improved execution time compared to our synthetic benchmark. This can be seen with the 1D Stencil distributed scenario (see Figure~\ref{fig:1d_stencil_distributed_errors}).

For replicate variations, the overheads remain similar to the one without faults. We do not observe contention as there is always sufficient work to be done during the program execution. We recommend resilience distributed APIs to invoke critical, highly parallel, low data input based functions. Invoking a single function on another locality that does trivial computation is potentially better off using local resilience variants, unless the output is of prime importance.

\subsection{1D stencil Local}
\label{subsection:results:1d_stencil_local}

We port 1D stencil to support resilience to check how our implementation performs on a real world application. For resilience replay, adding resilient local facilities to the base implementation adds minimal overheads. The additional time arises from the checksum computation and checksum validation and not the resiliency boilerplate code. For resilience replicate, the execution time is close to three times the base execution time, which is expected as it creates three replicas to work with. Overall, the execution times observed are well within the range we expected and shows that the implementation boilerplate code has no measurable impact on the execution time.

\begin{figure}[htbp]
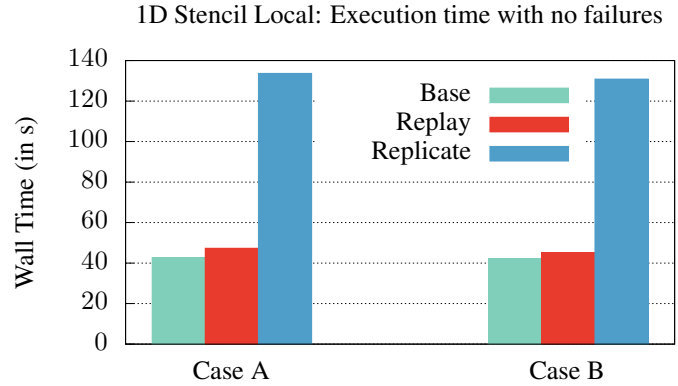

    \centering
    \begin{gnuplot}[terminal=epslatex]
    set terminal epslatex size 3.8in,2.25in
    set ylabel 'Wall Time (in s)'
    unset xlabel

    set yrange[0:140]

    set title '1D Stencil Local: Execution time with no failures'
    
    set key at 2.3,130
    
    set boxwidth 0.3
    set style fill solid
    
    set xtics ("Case A" 0.3, "Case B" 2.2)
    
    set grid ytics
    set ytics nomirror
    set xtics nomirror

    plot "plots/1d_stencil/overheads" every 3 using 1:2 with boxes title "Base" lt rgb "#80cfb9",\
     "" every 3::1 using 1:2 with boxes title "Replay" lt rgb "#e93b2d",\
     "" every 3::2 using 1:2 with boxes title "Replicate" lt rgb "#4f9ec9"
    \end{gnuplot}
    \caption{1D Stencil Local: Wall time for non-resilient base implementation and resilient variations. Case A works on 384 subdomains each with 8,000 data points. Case B works on 192 subdomains each with 16,000 data points. Both cases iterate over 4096 iterations with 256 time steps per iteration.}
    \label{fig:1d_stencil_overheads}
\end{figure}

\begin{figure}[htbp]
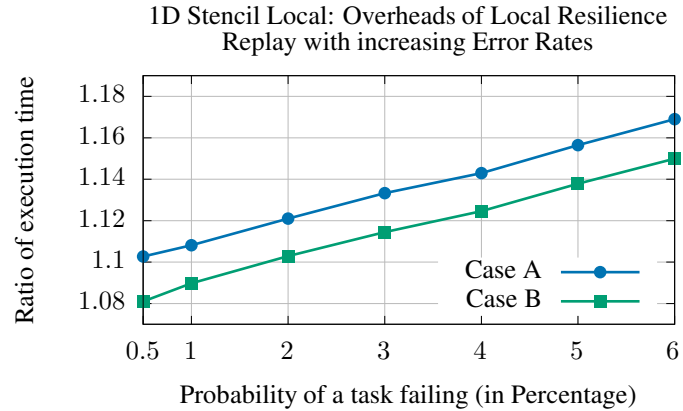

    \centering
    \begin{gnuplot}[terminal=epslatex]
        set terminal epslatex size 3.8in,2.25in
        set ylabel 'Ratio of execution time'
        set xlabel 'Probability of a task failing (in Percentage)'
        
        set grid ytics lt 2 lc rgb "#bbbbbb"
        set grid xtics lt 2 lc rgb "#bbbbbb"
        
        set title '\shortstack{1D Stencil Local: Overheads of Local Resilience\\Replay with increasing Error Rates}'
        set key bottom right
        
        set yrange[1.07:1.19]
        
        set xtics (0.5,1,2,3,4,5,6)

        plot \
        'plots/1d_stencil/errors_384' using 1:($3/$2) title 'Case A' lw 4 lt 7 lc 6 ps 1.5 w lp, \
        'plots/1d_stencil/errors_192' using 1:($3/$2) title 'Case B' lw 4 lt 5 lc 2 ps 1.5 w lp 
    \end{gnuplot}
    \caption{Ratio of execution times for resiliency replay validate to base non-resilient API. Case A works on 384 subdomains each with 8,000 data points. Case B works on 192 subdomains each with 16,000 data points. Both cases iterate over 4096 iterations with 256 time steps per iteration.}
    \label{fig:1d_stencil_errors}
\end{figure}

Figure~\ref{fig:1d_stencil_errors} shows the execution time as the number of failing tasks increases. Injecting errors within 1D stencil shows similar trends as seen in synthetic local workloads. For low failure rates, we observe that the overheads are about the same as the implementation overheads. As expected we see a spike in overheads as the number of failures increases. The replicate variations works independent of the percentage errors, so the execution time remains the same as base resilience replicate reported in Figure~\ref{fig:1d_stencil_overheads}.

We chose 1D stencil with local and distributed resilience APIs as stencil codes work on large grid sizes. Transferring the whole grid over the wire to enable distributed resilience will lead to a multi-fold slowdown. This is because an iteration of stencil on modern processors takes up to a few milliseconds. Transferring a large grid to another locality will take more time than the time taken to complete the iteration itself. Furthermore, the resultant grid needs to be transferred back to the locality invoking it. During the communication period, processors are bound to starve causing a noticeable slowdown. Thus, using 1D stencil as a benchmark allows us to discuss alternative ways to cater to such situations. We recommend using either local resilience variations for such applications, or a mix of local and distributed resilience as we showcase with the distributed 1D stencil benchmark, or invoking multiple iterations as a single task using distributed resilience variations. 

\subsection{1D stencil Distributed}

To test our distributed resilience performance, we extend the 1D stencil local version to support distributed scenario as described in ~\ref{subsection:stencil_distributed}. To the distributed 1D stencil application, we add checksum based resilience. To achieve the best performance, we use a mix of local and distributed resilience. An iteration is first tried up to three times locally. Upon failing to return a valid result, we switch to distributed resilience. Furthermore, the input action to the distributed resilience utilizes the local resilience facility to minimize further network traffic. This approach minimizes communication overheads while ensuring a valid output is returned.

\begin{figure}[tb]
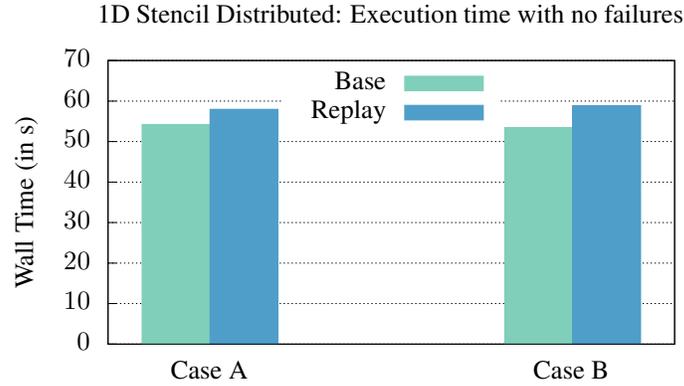

    \centering
    \begin{gnuplot}[terminal=epslatex]
    set terminal epslatex size 3.8in,2.25in
    set ylabel 'Wall Time (in s)'
    unset xlabel

    set title '1D Stencil Distributed: Execution time with no failures'
    
    set yrange [0:70]
    
    set boxwidth 0.3
    set style fill solid
    
    set xtics ("Case A" 0.15, "Case B" 1.75)
    
    set key top center
    
    set grid ytics
    set ytics nomirror
    set xtics nomirror

    plot "plots/1d_stencil_distributed/overheads" every 2 using 1:2 with boxes title "Base" lt rgb "#80cfb9",\
     "" every 2::1 using 1:2 with boxes title "Replay" lt rgb "#4f9ec9",\
    \end{gnuplot}
    \caption{1D Stencil Distributed: Wall time for non-resilient base implementation and resilient variations. Case A works on 384 subdomains each with 8,000 data points. Case B works on 192 subdomains each with 16,000 data points. Both cases iterate over 512 iterations with 2048 time steps per iteration.}
    \label{fig:1d_stencil_distributed_overheads}
\end{figure}

Figure~\ref{fig:1d_stencil_distributed_overheads} shows the execution time for non-resilient and resilient 1D stencil applications. Utilizing a large number of steps per iteration allows us to hide network latencies while the data moves through the wire. The base implementation takes about 10s more than the 1D stencil optimized for single node. The difference arises mostly due to two reasons; first, the network traffic and the additional distributed facilities required for implementation; second, the decision to choose a lockstep based implementation as opposed to a dataflow based implementation. Adding distributed resilience adds similar overheads as observed for 1D stencil local, mostly due to the additional checksum facilities.

Adding faulty nodes (see ~\ref{subsection:errors}) does not increase the overall execution time. This is because our 1D stencil application is implemented in an iteration lockstep manner. This means that the next iteration does not begin until the previous iteration completes. A faulty node in general does more replays and can lead to distributed resilience invocation. This means that the execution time depends directly on critical path drawn by the faulty node. Adding a single faulty node leads to deficient localities where the iteration has successfully executed. Therefore, increasing the number of faulty nodes does not lead to any visible increase in execution times as other localities take up the job to complete the iteration step. Furthermore, we do not see significant rise in execution times like we saw with the synthetic distributed workload. This is because we manually load balance the algorithm to ensure that an action is invoked not just on the next locality by the rank but on several localities to even out the imbalance.

As discussed in ~\ref{subsection:results:1d_stencil_local}, using purely distributed resilience facilities to achieve resilience would be detrimental to the execution times. One can see significant slowdowns as the data is transferred over the network.

\section{Conclusion}
In this paper, we discuss AMT resilience for both local only and distributed applications. We discuss the design choices taken and prototype resilience APIs based on those design choices in HPX. We implement two classes of resilience, namely task replay and task replicate. Task replay reschedules a task up to n-times until a valid output is returned. Task replication runs a task n-times concurrently. We demonstrate that only minimal overheads are incurred when utilizing these resiliency features for real world work loads.

We discuss how applications that work on large sized data can be detrimental to the performance of distributed resiliency due to network bottlenecks. We then describe ways to implement benchmarks such that these bottlenecks are mitigated. The paper also lays out ways to mix local and distributed resilience APIs in real world benchmarks.

Furthermore, as the new APIs are designed such 
that they are fully conforming to the C++ standard, these features will be easy enough to 
embrace and enable a seamless migration of existing code. 
Porting a non resilient 
application to its resilient counterpart requires minimal changes, 
along with the implementation of validation/vote functions, wherever necessary. 
This removes the necessity of costly code re-writes as well 
as time spent learning new APIs.

\section*{Future Work}

The proposed work is currently limited to soft failures that are detected with user provided error detections. The design choices for implementing distributed resilience help us to extend the facilities to hard faults, i.e., situations where a node goes offline. Furthermore, the distributed resilience needs be optimized to cater to non-uniform workloads. This can be done by attaching a load balancer to the resilience facilities to replay a task on a node with starving processor. Distributed replicate facilities can be further optimized by introducing the concept of cancellable actions. Using this concept, we will be able to terminate a remotely running action if a valid result is obtained from one of the replicates. This should bring down the execution time of non-uniform workloads significantly.

\section*{Acknowledgment}

Sandia National Laboratories is a multimission laboratory managed and operated by National Technology \& Engineering Solutions of Sandia, LLC, a wholly owned subsidiary of Honeywell International Inc., for the U.S. Department of Energy's National Nuclear Security Administration (NNSA) under contract DE-NA0003525. This work was partially funded by NNSA's Advanced Simulation and Computing (ASC) Program. This paper describes objective technical results and analysis. Any subjective views or opinions that might be expressed in the paper do not necessarily represent the views of the U.S. Department of Energy or the United States Government.

\bibliographystyle{unsrt}
\bibliography{references}

\end{document}